# Issues about the Adoption of Formal Methods for Dependable Composition of Web Services

Manuel Mazzara, Michele Ciavotta

Politecnico di Milano, Italy

mazzara@elet.polimi.it, ciavotta@elet.polimi.it

**ABSTRACT**

Web Services provide interoperable mechanisms for describing, locating and invoking services over the Internet; composition further enables to build complex services out of simpler ones for complex B2B applications. While current studies on these topics are mostly focused - from the technical viewpoint - on standards and protocols, this paper investigates the adoption of formal methods, especially for composition. We logically classify and analyze three different (but interconnected) kinds of important issues towards this goal, namely foundations, verification and extensions. The aim of this work is to individuate the proper questions on the adoption of formal methods for dependable composition of Web Services, not necessarily to find the optimal answers. Nevertheless, we still try to propose some tentative answers based on our proposal for a composition calculus, which we hope can animate a proper discussion.

**1 BACKGROUND**

Service Oriented Architecture and the related paradigm are modern attempts to cope with old problems connected to B2B and information interchange. Many implementations of this paradigm are possible but presently the so called Web Services look to be the most prominent, mainly because the underlying architecture is already there; it is simply the web which has been extensively used in the last 15 years. We can easily exploit HTTP (W3C.org), XML (W3C.org), SOAP (Box et al., 2000) and WSDL (Christensen et al., 2001). The World Wide Web provides a perfect basic platform to connect different companies and customers but cannot fulfill all the needs that use to arise in this context. It is perfect for the interconnection on a point-to point basis, but one of the B2B complication is the management of causal interaction between different services and the way in which the messages between them need to be handled, not always in a sequential way for example. This area of investigation is called composition, i.e. the way to build complex services out of simpler ones (Peltz, 2003). So, the need for workflow technology is evident. The positive thing is that we had this technology investigated for decades and we have also excellent modeling tools providing verification features that are grounded in the very active field of concurrency theory research.

Different organizations are working on composition proposals. The most important in the past have been IBM's WSFL (Leymann, 2001) and Microsoft's XLANG (Thatte, 2001). These two have then converged in Web Services Business Process Execution Language (Jordan & Evdemon) (WS-BPEL or BPEL for short), which is presently an OASIS Standard. The language allows workflow-based composition of services and in the committee members words the aim is

*Enabling users to describe business process activities as Web services and define how they can be connected to accomplish specific tasks.*

Earlier versions of the language were not so clear, the specification was huge and many points not very clear, especially in relation to the recovery framework and the interactions between different mechanisms (fault handlers and compensation handlers). The sophisticated implicit mechanism of recovery was confused. However, in the final version of the specification (which is lighter and clearer) fault handling during compensation has been clarified. WS-BPEL represents a necessary business tradeoff where not all the single technical choices have been done considered the entire set of options. For this reason, critics to the specification may lead to further improvement. In the following, we present a few questions that could help us in this process. We tried to logically separate these questions in three areas: Foundational questions, Verification questions and Extensions questions. For each area you can find a dedicated section.

## 2 FOUNDATIONS

The need for formal foundation has been discussed in recent last years and several attempts of using some kind of formal methods in this setting have been speculated. Some communities, for example, criticized the process algebra option (van der Aalst, 2003) promoting the Petri nets choice. Do we really need a formal foundation and which kind of formalism do we need? It is crucial to understand the notion of killer applications and to identify a possible selling point for our work. Furthermore, having worked for some time on Semantic Web Services, we are still trying to figure out whether adding a semantic description of services may bring a significant value in comparison with analysis and development costs spent over the last few years. So it is worth spending some time to investigate this issue as well. This section discusses more broadly these general and foundational issues.

## 2.1 Do we need formal foundations?

Functional programming languages have a formal foundation in the lambda-calculus. In Benjamin Pierce words:

*The lambda-calculus holds an enviable position: it is recognized as embodying, in miniature, all of the essential features of functional computation. Moreover, other foundations for functional computation, such as Turing machines, have exactly the same expressive power. The inevitability of the lambda-calculus arises from the fact that the only way to observe a functional computation is to watch which output values it yields when presented with different input values.*

The pi-calculus is a theory of mobile systems, which provides a conceptual framework for understanding mobility and mathematical tools for expressing mobile systems and reasoning about their behaviors. It introduces mobility generalizing the channel-based communication of CCS by allowing channels to be passed as data through rendezvous over other channels. In other words, it is a model for prescribing (specification) and describing (analysis) concurrent systems consisting of agents, which mutually interact and in which the communication structure can dynamically evolve during the execution of processes. Here, a communication topology is intended as the linkage between processes, which indicates which processes can communicate with which. Thus, changing the communication links amounts to a processes moving inside this abstract space of linked processes.

The symmetry between lambda-calculus and the pi-calculus could suggest some analogies. The option of building concurrent languages (and workflow languages too) on a formal basis has been investigated in several papers so far, also in connection with WS-BPEL. However, formal methods should bring mathematical precision to the development of computer systems providing precise notation in specifications and verification in design but so far WS-BPEL has not yet really been proved in an interesting relation with process algebras and we do not have conceptual tools for analysis, reasoning and software verification. Without providing this, all the hype about mathematical rigor is just pointless. Furthermore, as already said, a critical point in the BPEL specification was the definition of the recovery framework, which is actually critical for deploying dependable, composed web services.

webpi$\infty$ (Mazzara & Govoni, 2005) has been introduced to investigate how to use process algebra as a foundation in this context. It is a simple and conservative extension of the pi-calculus where the original algebra is augmented with an operator for asynchronous events raising and catching in order to enable the programming of widely accepted error handling techniques (such as long running transactions and compensations) with a reasonable simplicity. We addressed the problem of composing services starting directly from the pi-calculus and considering our proposals as foundational models for composition simply to verify statements regarding any mathematical foundations of composition languages and not to say that the pi-calculus is more suitable than other models (such as Petri nets) for these purposes

## 2.2 Which kind of killer application are we looking for?

Firstly, what is a killer application? Usually for killer application we intend any desirable computer program that can provide the core value for a technology. A killer application should increase sales for the underlying technology. The Wikipedia definition is: "A killer app is an application so compelling that someone will buy the hardware or software components necessary to run it"'. Then, what can we build so appealing that people would buy the heavy and complicated mathematical framework behind? We have also to discuss killer applications in term of performance and reliability (the "meta-level" explained above). Many of the theories we to use to build our tools, composition engines in this case but more generally any kind of tools, need to be justified in terms of costs/benefits.

Something that could be considered as a touchstone killer application is the Amazon Web Services (AWS), which are a collection of web services offered over the Internet by Amazon.com. Amazon Web Services can be accessed via HTTP, using SOAP. Amazon claims that, at June 2007, more than 330,000 developers had signed up to use the services. Anyway, it is not clear how the composition of services and the verification of properties can here play their roles. What is sure is that the wide use of AWS could be considered for targeting intensive experiments on this platform. Once we got some result in this scenario could be much easier to sell them in the scientific community and in the industrial world.

## 2.3 Semantic or syntactic approach to composition?

With the goal of giving technological support for the service-programming model, different approaches have been developed. The main classification that we can do is to distinguish between the syntactic and semantic approaches. The syntactical approach presently finds its main advantage in having concrete and easy to use compositional tools, while the semantics one should add semantics to web services standards (Sivashanmugam et al., 2003), especially the main advantages from service annotation and semantic discovery.

From the industry point of view, the syntactic approach has been largely understood and accepted, while the semantic one, although it promises interesting developments, is still lacking some concrete supports. With the syntactic approach the interface of a service is defined in an Interface Definition Language called WSDL, which is very close do the CORBA IDL in some sense. Basically the service is seen like an RPC with the relative signature (i.e. the syntax of messages entering and leaving the service). The order of messages exchange between the services is instead defined with other languages, e.g. WS-BPEL or similar. The approach lacks of a semantic definition of components, since WSDL is indeed only a syntactical interface definition.

The semantic approach finds its root in a different community, the Semantic Web community. Semantic Web services are the fruitful combination of Semantic Web and Web service technologies. The purpose of Semantic Web services is to overcome the limitations of current Web services by adding explicit semantics to them. The exploitation of such semantics would then resolve the interoperability issues and automate the Web service discovery and usage process. OWL-S (OWL Service Ontology) (http://www.w3.org/2004/OWL/) looks to be the major initiatives for providing semantic annotations on top of a Web Service infrastructure, by means of the three core ontologies: service profile, service model and grounding. Thereby, service profile presents "what a service does", service model describes "how a service works" and finally grounding supports "how to access it". SAWSDL (Semantic Annotations for WSDL and XML Schema) (http://www.w3.org/2002/ws/sawsdl/), the extension of WSDL-S, is the other recommended specification by W3C to provide more semantic annotation mechanisms to disambiguate the description of Web Services during automatic discovery and composition.

The two approaches have someway also different goals and different underlying formalisms. To our knowledge, semantics can be useful for service discovery, but has less to do with the way in which we can cope with dependability (and partly with composition in general), because it is not very clear which would be the concrete and ultimate advantage. For this reason, although we have investigated some solutions in the semantic web setting (Yan et al., 2007), hereby we prefer to spend effort especially on syntax since this community is progressing fast. The question regarding the adoption of semantic technologies remains still open anyway.

There are also some "hybrid solutions" where BPEL uses some "semantic discovery service" and discovery and matchmaking are performed by querying some knowledge base. Integration of semantic web services technologies into Business Process is indeed possible and Service Oriented applications can leverage advantages offered by these technologies. For this purpose, we proposed BPMO (Business Process Modeling Ontology) (Yan et al., 2007) and tried to ground it into standardized BPEL as the final implementation. BPEL lacks proper support for generating dynamic compositions and it does not really support explorative (on the fly) orchestration. In fact, we can say, (standard) BPEL has a static process composition where partner's discovery and bounding at run time is not possible. Although there are these proposals to complement BPEL with dynamic binding capabilities, so far only the implementations behind partner's services can change, not really the "interface".

# 3 VERIFICATION

Another interesting issue is, once we decided that formal methods can bring interesting advantages in developing composition languages, how this can be a matter of research in dependability? Why do we care about dependability in service composition? What does it mean in this setting and how can it be achieved? What can formal methods bring to the state-of-the-art? Furthermore, for verification purposes, which kind of software/conceptual tools are we interested in for specifying and verifying systems? And, in turn, which properties these verification tools should satisfy? In this section we try to gain a deeper insight into these issues.

## 3.1 How can we reach dependability?

Dependability in WS-standards applies only where SOAP is employed as an XML at message level, that means at the messaging protocol level (SOAP is not compulsory in SOA, though).

– WS-Reliability (OASIS): it adds dependability to the unreliable communication channel of the Internet
– WS-Security (OASIS): it specifies mechanisms to provide integrity and confidentiality in SOAP messages

However, things are more complex since loosely coupled components like Web services, being autonomous in their decisions, may refuse requests or suspend their functionality without notice, thus making their behavior unreliable to other activities. For this reason, most of the web languages also include the notion of loosely coupled transaction – called as *web transaction* (Little, 2002) in the following – defined as a unit of work involving loosely coupled activities that may last long periods of time. These transactions, being orthogonal to administrative domains, have the atomicity and isolation properties relaxed and, instead of assuming a perfect rollback in case of failure, they support the explicit programming of compensation activities. Web transactions usually contain the description of three processes: body, failure handler, and compensation. The failure handler is responsible for reacting to events that occur during the execution of the body. Hence, when those events occur, the body is blocked and the failure handler is activated. The compensation, on the contrary, is installed when the body commits; it remains available for outer transactions requiring the undo of previously performed actions. BPEL also uses this approach.

Dependable composition is not standardized at all, as far as we know. This topic can be categorized as follows:

– **Fault prevention**: it can be performed at level of single services by domain-specific techniques. Oracle BPEL process manager/Biztalk provide indeed an initial support

– **Fault removal**: verification via Static Analysis includes contract conformance and deadlock safety. Only a few works on these topics appeared in literature
– **Fault forecasting**: we think it is a not approached issue. Maybe stochastic Petri nets could help
– **Fault tolerance**: we focused on recovery and we will try to give a brief introduction in this paper

Our recovery approach is described in (Lucchi & Mazzara, 2007) where we showed that different mechanisms for error handling are not necessary and presented the BPEL semantics in terms of webpi∞, which is based on the idea of event notification as the sole error handling mechanism. This result allows us to extend any semantic considerations about webpi∞ to BPEL. Other papers discussing the formal semantics of compensable activities in this context are: the work by Hoare (Hoare, 2002), which is mainly inspired by XLANG, the calculus of Butler and Ferreira (Butler & Ferreira, 2007), which is inspired by BPBeans, the pit-calculus (Bocchi et al., 2003) considering BizTalk and (Bruni et al., 2002) dealing with short-lived transactions in BizTalk. The work in (Bruni et al., 2005) also presents formal semantics for a hierarchy of transactional calculi with increasing expressiveness.

### 3.2 Which kind of verification tools do we need?

In the case of BPEL, any verification on compositions has to check that the basic services can work together by means of opportune interactions. Although verification is not strictly required for the execution, performing state-of-the-art static analysis on processes (e.g. deadlock-freedom analysis and useless-code elimination) can be worthwhile for designers. No tool can be complete for theoretical reasons but they could be really useful in concrete cases.

Different formal methods provide specific advantages in this sense. The pi-calculus, as discussed, is profitable in modeling mobility and interactions. This looks certainly appealing given the current developing trend of business processes. If the final goal is to use the pi-calculus for modeling and verification of workflows, like in BPEL, we should be able to identify the specific purpose of workflow verification.

In literature a few approaches have been presenting leveraging the pi-calculus (or CCS) and they can be synthesized as follows:

– *Deadlock*: a deadlock refers to a situation in which a workflow instance gets into a state such that no more activities can be executed. Kobayashi has explored deadlock freedom by means of typed process calculus. Tool support exists for his investigation (http://www.kb.ecei.tohoku.ac.jp/ koba/typical/).
– *Contract conformance*: the definition of a formal contract language for describing interactions of clients with Web services has been investigated in (Carpineti et al., 2006). They define a precise notion of compatibility between services, called subcontract relation, so that equivalent services can be safely replaced with each other.

A foundational unifying framework based on the pi-calculus that could be applied in this area has been developed in (Lucchi, & Mazzara, 2007) and (Mazzara, & Lanese, 2006). It is an orchestration language able to meet composition requirements and to encode the whole BPEL. Ultimately, these works together presents a powerful and expressive language, with a solid semantics, that allows formal reasoning and processes equivalence proofing. These results can be used for a better understanding of the BPEL semantics and behavior, in general misunderstood, and for developing effective tools to detect process equivalences leading to flow design simplification and orchestration engines and compilers lightening. With tools we intend both software and conceptual tools, namely methods for reasoning about programs.

Another approach for the specification of systems with forms of concurrency and interferences is described in (Coleman, & Jones, 2007) and (Collette, & Jones, 2000). These work are based on the notion of rely/guarantee rules. The specification of composed systems may be possible via rely/guarantee at the level of a single service. However, it is not evident how this would apply to the whole system, especially for recovery requirements. Probably, some formal definition of a weak (domain specific) form of consistency is required at this stage. Exception handling is the most general means for achieving application-specific recovery. In webpi$\infty$ we called it *event handling* since we did not want to commit on the term "exception" - we wanted to encode both compensations and exceptions into one (unification), so we needed a new term. The general idea of webpi$\infty$ is that we are offering programmers too many techniques for recovery, which looks too complex. This would call for a cleaning up, especially for what concern BPEL.

### 3.3 Which properties verification tools should satisfy?

The origin of the modern notion of formal methods could be grounded back to the design of the first compilers in some sense. Computer scientists at that time recognized that it is crucial to ensure the "correctness"' of the compiler since all the other programs would have been then subjected to the consequences of the compilation phases. How can you guarantee any kind of correctness (howsoever you want to define it) if you cannot state that the compiler works according to the specification, i.e. that indeed it does what it is supposed to do? To figure out how critical was the design of a compiler at the origin of high level programming languages, consider what could have happened if the designers of the first compiler would have inserted some kind of replicating trojan or virus. Such a malicious code would have propagated in every compiled program and so also in all the following compilers compiled with the first one. This would have been a serious attempt to mine the correctness of all the compiled programs but even if some naïve mistakes would have been introduced they could have caused many significant problems. We think the same is happening now when we work on verification tools for the desired properties. We believe the importance of this point is not adequately recognized nowadays. The question is: which kind of properties verification tools should satisfy? How can we define and verify those tools before using them on our applications? Which is the notion of correctness for these tools? How can we formally specify it and how formal methods can be applied at this "meta-level"?

**4 EXTENSIONS**

Original CCS provides a simple and clear way to define basic concepts of workflow. Why do not we use it for our purposes? It is of crucial importance to understand why so many researchers in recent years moved towards different formalisms, first of all the pi-calculus. There has been a proliferation of timed process algebra as well and several attempts of using them to model business processes. Instead, fault forecasting has been almost ignored by researchers. Could stochastic extensions for process algebra be useful in this situation? This section is dedicated to those language extensions that look promising in answering these questions.

**4.1 Do we need mobility?**

Although many papers use the term pi-calculus and process algebra interchangeably, there is a difference between them. Algebra is a mathematical structure with a set of values and a set of operations on the values. These operations have algebraic properties such as commutativity, associativity, idempotency, and distributivity. In typical process algebra, processes are values and parallel composition is defined to be a commutative and associative operation on processes.
The pi-calculus is a special algebra that differs from other models for concurrency, precisely for the fact that include a notion of mobility, i.e. some sort of dynamic reconfiguration. The pi-calculus looks interesting because of the way it deals with component bindings as first class objects, which enables this dynamic reconfiguration to be expressed simply. So, the question now is do we need this additional feature of the pi-calculus or should we restrict our choice to model, like CCS, without this notion of mobility? Why all this hype over the pi-calculus and a so rare focus on its crucial characteristic?

However, it seems that supporting link-passing mobility is an essential feature that composition languages should have. Indeed, while in some scenarios services can be selected at design-time, in others some services they can only be selected at runtime and this selection has then to be propagated to different parties. This phenomenon is called link-passing mobility and it is properly approached in (Decker et al., 2007).

**4.2 Do we need to model time?**

In a previous work (Mazzara, 2005), we addressed the notion of time as we recognized the limits of those works were time is not considered and the usefulness of time handling in programming business transactions. So we considered timed transactions, i.e. transactions that can be interrupted by a timeout. Real workflow languages presently provide this feature: XLANG, for instance, includes a notion of timed transaction as a special case of long running activity. BPEL also allows similar behaviors by means of alarm clocks. To meet the challenge of time in composition, webpi has been equipped with an explicit mechanism for time elapsing and timeout handling (Mazzara, 2005). By adding time it is possible to express more meaningful and realistic scenarios in composition. Berger-Honda Timed-pi (Berger, 2002) (Berger, & Honda, 2000), directly inspired webpi model of time, just the idle rule has been removed and some other minor variations applied. A synopsis between the two approaches, underlying differences and similarities, has been shown in (Mazzara, 2005).

Although the challenge of coping with timed process algebra was fascinating and timed transactions are something that can really be useful in practical scenarios, we think that introducing time into the model is for SOA an unnecessary complication, at least at the pi-calculus level and integrating everything in the same formalism. Modeling B2B presents features like process interaction, causality between messages, channel mobility (i.e. channels that can be discovered only at runtime) in a scenario where the partners own to different administrative domains. This requires the underlying framework to be simple. Timed transaction is an accessory that, although useful, does not represent the computational core (if we want to call this way) of the story. Adding time is not for free, it complicates the semantics, and it forces us to consider different models of time with the goal of modeling a feature which, in BPEL, is not even a first class citizen and it is instead obtained by the use of alarm clocks. This does not mean that time should not be modeled, but it has just to be considered at a different level. Similar considerations hold for stochasticity as we will discuss in the following.

**4.3 Do we need stochastic extensions?**

The discussion here shares similarities with time modeling as described in 4.2. The concrete difference is that, so far, we never faced stochastic extensions while we approached timed extensions. We have already introduced the dependability issue in this work: one of the most overlooked issues in service composition is fault forecasting. Fault forecasting is used to predict potential faults and their consequences for a specific system. Predictions can be qualitative or quantitative. When the evaluation is quantitative we apply mathematical concepts of probability theory to potential fault occurrences to get a precise measure. We believe it is crucial to consider dependability aspects in a quantitative manner exploiting instruments allowing us to describe random phenomena like spontaneous crashes.

Considering the wide acceptance of process algebras many Markovian extensions has been presented to cope with performance issues, generating a new research directions between concurrency and performance communities (Hillston, 2005), (Hermanns et al., 2002) All these extensions want to approach performance issues inside the model itself, at the same level. This, as it was for time, adds complications in modeling, semantics, etc. The open question here is if can be worth or not following this approach or if it would be the case of separating the concerns facing them on different layers. Our feeling is that separation brings simplicity and would allow a more accurate analysis of different, maybe disconnected, aspects.

## 5 THE COMPOSITION CALCULUS

In this section we present a proposal to cope with the issues presented above. Although webpi is ambitious, certainly we do not pretend to solve all the problems and to give the ultimate answer to all those questions. This paper is about the way in which webpi can be considered in the overall scenario of formal methods for dependable Web Services. Giving all the details about the language and its theory is far beyond the scope of this paper and would not fit page constraints. You can find all the relevant details in some previous work, especially in (Mazzara, 2006), (Lucchi, & Mazzara, 2007) and (Mazzara, & Lanese, 2006). Here we only recall the main concepts for the purpose of this paper, i.e. trying to give some tentative answers to the above questions. After presenting the language we will recall the above questions trying to readdress them using webpi.

### 5.1 Syntax

The syntax of webpi∞ processes relies on countable sets of names, ranged over by x, y, z, u, . . . We intend $i \in I$ with I a finite non-empty set of indexes.

$$
\begin{aligned}
P ::= \quad & 0 & &\text{(nil)} \\
| \quad & \overline{x} \langle \widetilde{u} \rangle & &\text{(output)} \\
| \quad & \sum_{i \in I} x_i(\widetilde{u_i}).P_i & &\text{(alternative composition)} \\
| \quad & (x)P & &\text{(restriction)} \\
| \quad & P \,|\, P & &\text{(parallel composition)} \\
| \quad & !x(\widetilde{u}).P & &\text{(guarded replication)} \\
| \quad & (\!| P \,;\, P |\!)_x & &\text{(workunit)}
\end{aligned}
$$

A process can be the inert process, an output, an alternative composition consisting of input guarded processes, a restriction, a parallel composition of processes, a replicated input or a workunit that behaves as the body P until an abort x is signaled and then behaves as the event handler Q. Names in outputs, inputs, and replicated inputs are called subjects of outputs, inputs, and replicated inputs, respectively. It is worth to notice that the syntax of webpi∞ processes simply augments the asynchronous pi-calculus with workunit process.

## 5.2 Semantics

We give the semantics for the language in two steps, following the approach of Milner (Milner, 1992), separating the laws that govern the static relations between processes from the laws that rule their interactions. The first step is to define a static structural congruence relation over syntactic processes. A structural congruence relation for processes equates all agents we do not want to distinguish. It is introduced as a small collection of axioms that allow minor manipulation on the processes' structure. This relation is intended to express some intrinsic meanings of the operators, as for example the fact that the parallel operator is commutative. The second step is to define, by means of an operational semantics, the way in which processes dynamically evolve. This way we simplify the statement of the semantics just closing with respect to $\equiv$, i.e., closing under process order manipulation induced by structural congruence.

**Definition 1.** *The* structural congruence $\equiv$ *is the least congruence satisfying the abelian monoid laws for parallel and summation (associativity, commutativity and* **0** *as identity) closed with respect to $\alpha$-renaming and the following axioms:*

1. *Scope laws:*

$$(u)\mathbf{0} \equiv \mathbf{0}, \quad (u)(v)P \equiv (v)(u)P,$$
$$P \,|\, (u)Q \equiv (u)(P \,|\, Q), \quad \text{if } u \notin \mathtt{fn}(P)$$
$$\langle\!(z)P \,;\, Q\rangle\!_x \equiv (z)\langle\!P \,;\, Q\rangle\!_x, \quad \text{if } z \notin \{x\} \cup \mathtt{fn}(Q)$$

2. *Workunit laws:*

$$\langle\!\mathbf{0} \,;\, Q\rangle\!_x \equiv \mathbf{0}$$
$$\langle\!\langle\!P \,;\, Q\rangle\!_y \,|\, R \,;\, R'\rangle\!_x \equiv \langle\!P \,;\, Q\rangle\!_y \,|\, \langle\!R \,;\, R'\rangle\!_x$$

3. *Floating law:*

$$\langle\!\overline{z}\,\langle\widetilde{u}\rangle \,|\, P \,;\, Q\rangle\!_x \equiv \overline{z}\,\langle\widetilde{u}\rangle \,|\, \langle\!P \,;\, Q\rangle\!_x$$

The scope laws are standard while workunit and floating laws are new. The first law defines committed workunit, namely workunit with 0 as body. Workunits of this form are semantically equivalent to 0 and, therefore, cannot fail anymore. The second law moves workunit outside parents, thus flattening the nesting. Notwithstanding this flattening, parent workunits may still affect children ones by means of names. The last law allows output messages to be moved outside workunit boundaries. By this law, messages are particles that independently move towards their inputs. The intended semantics is the following: if a process emits a message, this message traverses the surrounding workunit boundaries until it reaches the corresponding input. In case an outer workunit fails, recoveries for this message may be detailed inside the handler processes.

The reduction relation defines the dynamic behavior of processes, where we use the shortcut:

$$\langle\!| P\ ;\ Q |\!\rangle \stackrel{def}{=} (z)\langle\!| P\ ;\ Q |\!\rangle_z \text{ where } z \notin \texttt{fn}(P) \cup \texttt{fn}(Q)$$

**Definition 2.** *The reduction relation $\rightarrow$ is the least relation satisfying the following axioms and rules, and closed with respect to $\equiv$, $(x)\_$ , $\_ | \_$, and $\langle\!| \_\ ;\ R |\!\rangle_z$:*

$$\begin{array}{l}
(\textsc{com})\\
\quad \overline{x_i}\langle\widetilde{v}\rangle\ |\ \sum_{i \in I} x_i(\widetilde{u_i}).P_i \quad \rightarrow \quad P_i\{\widetilde{v}/\widetilde{u_i}\}\\
(\textsc{rep})\\
\quad \overline{x}\langle\widetilde{v}\rangle\ |\ !x(\widetilde{u}).P \quad \rightarrow \quad P\{\widetilde{v}/\widetilde{u}\}\ |\ !x(\widetilde{u}).P\\
(\textsc{fail})\\
\quad \overline{x}\langle\rangle\ |\ \langle\!|\ \prod_{i \in I} \sum_{s \in S} x_{is}(\widetilde{u_{is}}).P_{is}\ ;\ Q |\!\rangle_x \quad \rightarrow \quad \langle\!| Q\ ;\ \mathbf{0} |\!\rangle \quad (I \ne \emptyset)
\end{array}$$

Rules (com) and (rep) are standard in process calculi and models input-output interaction and lazy replication. Rule (fail) models workunit failures: when a unit abort (a message on a unit name) is emitted, the corresponding body is terminated and the handler activated. On the contrary, aborts are not possible if the transaction is already terminated (namely every thread in the body has completed its own work), for this reason we close the workunit by restricting its name. The reason for maintaining such a structure will be clear in the section relative to the labeled semantics.

You can find all the definitions and proofs with an extensive explanation for the extensional semantics, the notions of barb, process contexts and barbed bisimulation in (Mazzara, 2006). You can find also definitions for Labelled Semantics, asynchronous bisimulation, labeled bisimilarity and the proof that it is a congruence. There are also results relating barbed bisimulation and asynchronous labeled bisimulation and many examples. A core WS-BPEL is encoded in webpi∞ in (Mazzara, 2006) and some properties related to this encoding are proved for it.

### 5.3 Pragmatics

In this section we recall all the questions discussed in the first part of this paper, but in the light of the presented language. The discussion moves now towards a concept like pragmatic, i.e. practical usage of the ideas/notions/tools developed so far.

- *Do we need formal foundations?* webpi∞ starts from the assumptions that we do. Motivations are given in the proper section of this paper. The reader not convinced by those motivations would not find useful and interesting our proposal. The discussion cannot be pushed forward here for lack of space.

- *Which kind of killer application are we looking for?* We have begun working on webpi∞ starting form a simple case study that, in our opinion, contains all the "relevant logic" of real scenarios. The case study is described in (Mazzara, & Govoni, 2005). We will never emphasize enough how important is this idea of a killer app. It is about business and market, if we do not find any interesting killer app we are not going to get money and funding for any further research on the field. It is worth noting that we are not saying that theoretical models and math cannot have sense for themselves, indeed there are beautiful piece of math there make sense "as they are". Nevertheless, we think that, finding applications that can impact the people's life also on other levels than only contemplation of beauty is still possible also in this field.

- *Semantic or syntactic approach to composition?* The approach of webpi∞ is definitively on the syntax-side of the river. This does not mean that we exclude the other approach a priori; simply the target of our work is slightly different. We did not consider any semantic discovery or composition, although our use of channel mobility opens the possibility to provide some kind of dynamic "on the fly" (semantic supported) composition. We focused more on modeling static compositions where all the partners are known a priori, but channel mobility of webpi∞, inherited by its parent (the pi-calculus), does not preclude this option.

- *How can we reach dependability?* webpi∞ offers a unification of handlers and a simplification of the BPEL recovery system. As discussed in section 1 and 2, the complication of recovery and the many concepts involved does not simplify the designer/developers life. We believe our contribution is significant to shed light and reach a better understanding of WS-BPEL recovery framework to therefore simplify it.

- *Which kind of verification tools do we need?* Tools to enforce some form of correctness at design time must be considered as relevant for Web Services composition. Orchestration engine on the market so far provide graphical tools that help designer to create compositions that are well formed by construction but this support is not enough for important B2B applications. The theory behind webpi∞ allows the definition of behavioral equivalences, which are one way to check for certain properties, as discussed in section 5.3.

- *Which properties verification tools should satisfy?* Verification tools for webpi∞, considering the underlying theory, cannot be complete in the sense that, the desirable properties we discussed in this paper are not fully decidable. So we cannot expect to have the ultimate solution to the problem. Limitations of any related tool are grounded back in well-known decidability results. Finally, the correctness of the final implementations is surely a requirement that can be enforced with already known tools used in compiler industry and in software engineering.

- *Do we need mobility?* We believe is a relevant feature for B2B and webpi$\infty$ inherits it form the pi-calculus. Channel mobility is also important since, as we have already stated before, can be exploited to provide dynamic "on the fly" (semantic supported) composition. In (standard) BPEL all the partners are known a priori, but webpi$\infty$ is able take into account further evolutions of the language.

- *Do we need to model time?* There is a timed version of webpi$\infty$ that tries to cope with BPEL timed transactions. The point here is that timed process algebra is really complex since the introduction of time largely complicates the theory. We still have to convince ourselves that this worthwhile endeavor. More investigation is left as future work.

- *Do we need stochastic extensions?* This question is indeed out of the scope of webpi$\infty$, which was not intended for fault forecast purposes, for example. Any other investigation here is left as future work.

## 6 CONCLUSIVE THOUGHTS

The webpi$\infty$ development with its related theory has required a lot of effort. Moreover writing down and synthesizing in this paper all the issues emerged during these years has been somehow exciting but at the same time exhausting. The bigger question now really is: are the answers that webpi$\infty$ can give (at least partially) satisfactory to the many question discussed here? Can we have the feeling that it is the case to carry on in this direction? Should we investigate somewhere else, in some more promising direction? We hope with this work to have given the reader an overall view of this research area. We now expect further questions arising and a fruitful discussion taking off.

**Acknowledgments**

These issues have been discussed during the useful conversations with Cliff Jones and Ani Bhattacharyya. For some of the ideas discussed here we have also to thank Cosimo Laneve, and Gianluigi Zavattaro among the others.